\documentclass[%
 aip,
 amsmath,amssymb,
 reprint,%
]{revtex4-1}

\usepackage{graphicx}% Include figure files
\usepackage{dcolumn}% Align table columns on decimal point
\usepackage{bm}% bold math
\usepackage[utf8]{inputenc}
\usepackage[T1]{fontenc}
\usepackage{mathptmx}
\usepackage{etoolbox}

\makeatletter
\def\@email#1#2{%
 \endgroup
 \patchcmd{\titleblock@produce}
  {\frontmatter@RRAPformat}
  {\frontmatter@RRAPformat{\produce@RRAP{*#1\href{mailto:#2}{#2}}}\frontmatter@RRAPformat}
  {}{}
}%
\makeatother
\begin{document}

\preprint{AIP/123-QED}

\title[]{Preservation of scalar spin chirality across a metallic spacer in synthetic antiferromagnets with chiral interlayer interactions}
% Force line breaks with \\
\author{Miguel A. Cascales Sandoval*}
\affiliation{Institute of Applied Physics, TU Wien, Wiedner Hauptstra{\ss}e 8-10, Vienna, 1040, Austria}
\affiliation{SUPA, School of Physics and Astronomy, University of Glasgow, Glasgow, G12 8QQ, UK}%

\author{A. Hierro-Rodr{\'i}guez*}%
\email[Corresponding author e-mails: ]{miguel.cascales@tuwien.ac.at, hierroaurelio@uniovi.es, amalio.fernandez-pacheco@tuwien.ac.at.}
\affiliation{Departamento de F{\'i}sica, Universidad de Oviedo, Oviedo, 33007, Spain}
\affiliation{CINN (CSIC-Universidad de Oviedo), El Entrego, 33940, Spain}
\affiliation{SUPA, School of Physics and Astronomy, University of Glasgow, Glasgow, G12 8QQ, UK}

\author{S. Ruiz-G{\'o}mez}
\affiliation{Max Planck Institute for Chemical Physics of Solids, Dresden, 01187, Germany}

\author{L. Skoric}
\affiliation{University of Cambridge, Cambridge, CB3 0HE, UK}

\author{M. A. Niño}
\affiliation{ALBA Synchrotron Light Facility,
Cerdanyola del Vall{\'e}s, 08290, Spain}

\author{S. McVitie}
\affiliation{SUPA, School of Physics and Astronomy, University of Glasgow, Glasgow, G12 8QQ, UK}

\author{S. Flewett}
\affiliation{Deutsches Elektronen Synchrotron, Hamburg, 22607, Germany}
\affiliation{Center for Data and Computing in Natural Sciences, Hamburg, 22761, Germany}
%\affiliation{Instituto de F{\'i}sica, Pontificia Universidad Cat{\'o}lica de Valpara{\'i}so, Avenida Universidad 330, Valpara{\'i}so, Chile}

\author{M. Foerster}
\affiliation{ALBA Synchrotron Light Facility,
Cerdanyola del Vall{\'e}s, 08290, Spain}

\author{Elena Y. Vedmedenko}
\affiliation{Department of Physics, University of Hamburg, Hamburg, 22607, Germany}

\author{N. Jaouen}
\affiliation{SOLEIL Synchrotron, L'orme des Merisiers, 91192 Gif-Sur-Yvette, Cedex, France}
\affiliation{Department of Molecular Sciences and Nanosystems, Ca' Foscari University of Venice, Via Torino 155, Mestre, I-30172, Italy}

\author{A. Fern{\'a}ndez-Pacheco*}
\affiliation{Institute of Applied Physics, TU Wien, Wiedner Hauptstra{\ss}e 8-10, Vienna, 1040, Austria}


\begin{abstract}
Chiral magnetic textures are key for the development of modern spintronic devices. In multilayered thin films, these are typically stabilized via the interfacial intralayer Dzyaloshinskii-Moriya interaction (DMI). Additionally, it has been recently observed that DMI may also promote vector spin chirality along the third dimension, coupling spins in different magnetic layers via non-magnetic spacer layers, an effect referred to as interlayer DMI (IL-DMI). This interaction holds promise for 3D nanomagnetism, from the creation of 3D spin structures such as hopfions to new forms of magnetic functionality in the vertical direction via remote control of chiral spin states.
 
Here, we investigate via magnetic X-ray scattering and imaging techniques the chiral nature of orthogonal magnetic states that form in a synthetic antiferromagnet with IL-DMI. We find that the vector spin chirality of the textures formed in an in-plane layer is determined by the net out-of-plane spin configuration of a neighboring layer, leading as a result to complex spin states across a metallic interface where the overall scalar spin chirality is preserved. This work thus uncovers a new flavor of chiral interlayer interactions, demonstrating new ways to control magnetic chirality in three dimensions.\\

%This work thus uncovers a new flavor of chiral interlayer interactions which provide new ways to control magnetic chirality in three dimensions.
\end{abstract}

\maketitle

\section{Introduction}

Spin chirality \cite{ishizuka2020anomalous}, a key concept in modern spintronic applications \cite{yang2021chiral,naaman2015spintronics}, refers to the spatial rotation of spins following a well defined winding sense. Chirality is typically induced by the antisymmetric exchange or Dzyaloshinskii-Moriya interaction (DMI) \cite{moriya1960anisotropic}, which favors chiral orthogonal coupling between neighboring spins. This interaction is mathematically described by the atomistic Hamiltonian \cite{coey_2010} $\mathcal{H}_{\text{DMI}} = \vec{D}_{ij}\cdot(\vec{s}_{i}\times\vec{s}_{j})$, where $\vec{D}_{ij}$ is the DMI vector derived from the Moriya rules \cite{crepieux1998dzyaloshinsky}, and is dependent on the spatial positions of neighboring spins $\vec{s}_{i}$ and $\vec{s}_{j}$.

\begin{figure*}[ht]
\centering
\includegraphics[scale=0.133]{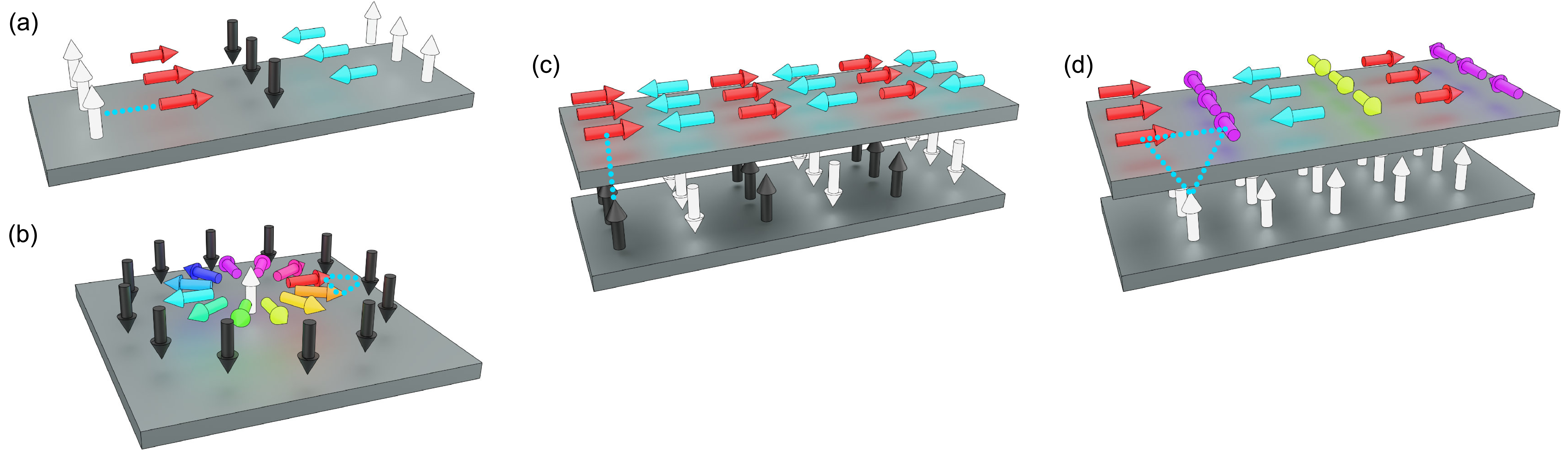}
\caption{\label{fig:figure1} Magnetic (a) chiral 360$^{\circ}$ DW and (b) magnetic skyrmion induced via intralayer DMI in individual ferromagnetic multilayers (heavy metal layers omitted in all sketches). (c) IL-DMI ground state configuration for two ferromagnetic layers with orthogonal anisotropy axes ("T-configuration"). (d) Simplified representation of magnetic state found in experiments, where the IP layer has a well defined vector chirality, and the OOP layer is fully saturated. The dashed lines connecting two spins denote the direction used to compute the vector chirality, whereas the lines connecting three spins represent scalar spin chirality. Colored arrows denote the 3D orientation of the magnetization vector according to the \textit{hsl} colormap.}
\end{figure*}

In ferromagnetic layers forming part of a multilayer, intralayer DMI couples spins within the film plane leading to the formation of chiral spin configurations with attractive properties for spintronics. This includes chiral domain walls (DWs) as shown in figure \ref{fig:figure1} (a) \cite{hoffmann2015opportunities,kim2018chirality,ajejas2017tuning}, and topologically non-trivial structures like skyrmions, figure \ref{fig:figure1} (b) \cite{tokura2020magnetic,kang2016skyrmion,hayami2022skyrmion}. These noncollinear textures possess vector spin chirality, computed as $\vec{s}_{i}\times\vec{s}_{j}$. However, for describing chiral systems with more intricate, noncoplanar spin configurations such as the skyrmion, scalar spin chirality $\vec{s}_{i}\cdot(\vec{s}_{j}\times\vec{s}_{k})$ is the more appropriate metric \cite{li2022all,fujiwara2021tuning,grytsiuk2020topological,ishizuka2018spin}. Magnetic configurations with non-zero scalar spin chirality are found to give rise to emergent magnetotransport effects \cite{grytsiuk2020topological}, \textit{e.g}, the topological Hall effect \cite{neubauer2009topological}.

Recently, it has been shown that a non-negligible DMI interaction may also exist along the third dimension, \textit{i.e.}, normal to the film plane. This interaction, referred to as interlayer DMI (IL-DMI) \cite{fernandez2019symmetry,vedmedenko2019interlayer,han2019long}, has been shown to promote a specific vector chirality in between spins in different layers, as sketched in figure \ref{fig:figure1} (c). IL-DMI originates from interlayer structural symmetry breaking in combination with the presence of non-magnetic atoms with large spin-orbit coupling (SOC).

So far, there are two models proposed to explain the IL-DMI. In the first model \cite{han2019long}, the effect manifests as an effective chiral field acting on the magnetization of each layer, caused by the chiral interaction with the spins of a neighboring layer. Ab-initio calculations indicate how microstructural changes in the stack, such as the shift of atoms in one layer with respect to their high-symmetry location would give such an effective field, with an in-plane DMI vector accounting for it. The second approach \cite{vedmedenko2019interlayer} is based on the generalization of the Levy-Fert atomistic model for DMI \cite{levy1981anisotropy} to multiple layers. For particular stack configurations or under moderate disorder, a nonzero interlayer DMI can appear, with a more complex arrangement of DMI vectors accounting for it. Whereas for the first case the ground state of the system consists of a uniform magnetic state for each layer, with an in-plane/out-of-plane orthogonal configuration following a well-defined sense of rotation for the magnetization from one layer to the next, the second predicts no IL-DMI energy gain for such a configuration. Instead, in the second model, IL-DMI is expected to play a role only when non-collinear spin states are present in at least one of the layers, with the chiral exchange bias being a consequence of non-uniform magnetic states during reversal \cite{fernandez2019symmetry}. So far, complex 3D spin states as those predicted by this model have not yet been observed.

Experimental works have been mostly focused on heavy metal/ferromagnetic systems, finding evidence of the IL-DMI as an effective chiral exchange bias which can be exploited, \textit{e.g.}, for field-free deterministic spin-orbit switching in perpendicularly magnetized multilayer films \cite{huang2022growth,wang2021spin}. This effective field also favors the stabilization of exotic but topologically-trivial spin textures such as 360$^{\circ}$ domain wall rings \cite{cascales2023rings}. The relationship between antisymmetric IL-DMI and symmetric RKKY interlayer interactions has been also experimentally established \cite{liang2023ruderman,li2024interlayer}. Furthermore, a recent work in a FM/Ag system reports large values for the IL-DMI, leading to the chiral reversal of the two in-plane magnetized layers, an effect which has been understood as a manifestation of IL-DMI via Rashba-SOC \cite{arregi2023large}. However, no direct experimental evidence of complex chiral spin states under the presence of IL-DMI have been reported so far. The microscopic nature of the interaction and symmetry of the DMI vectors are still to be fully established. 

Here, we perform experiments in a series of synthetic antiferromagnets (SAFs) with IL-DMI \cite{fernandez2019symmetry}, in order to investigate in detail whether chiral spin states form favored by this interlayer interaction. These SAFs have the magnetic layers in a "T-configuration", \textit{i.e.}, they consist of an in-plane (IP) and an out-of-plane (OOP) layer. We study the chiral nature of the textures forming in the SAF by a combination of X-ray resonant magnetic scattering (XRMS) and photoemission electron microscopy (PEEM). From these measurements we discover that OOP magnetic fields, which set the net orientation of the OOP layer, determine the sign of the vector spin chirality of the IP layer through interlayer interactions. These results show evidence that the overall interlayer scalar spin chirality of the multilayer is preserved across the non-magnetic spacer, as sketched in figure \ref{fig:figure1} (d). To complement our experiments, we discuss our results in terms of the two different approaches used to model IL-DMI, along with other possible mechanisms.

These results show evidence of complex chiral spin coupling in three dimensions between ferromagnetic layers separated by a metallic spacer, opening new oportunities in 3D spintronics.

%the formation of chiral IP textures upon demagnetizing the IP layer, whose sign of the vector chirality describing these states is directly controlled by the net orientation of the OOP layer, previously set by an OOP field saturating the SAF. In simpler terms,

%We find how all seem to fail in giving a clear explanation of these experimental results, therefore concluding that further development is necessary to fully describe the microscopic origin of the observed effect.

\begin{figure*}[ht]
\centering
\includegraphics[scale=0.15]{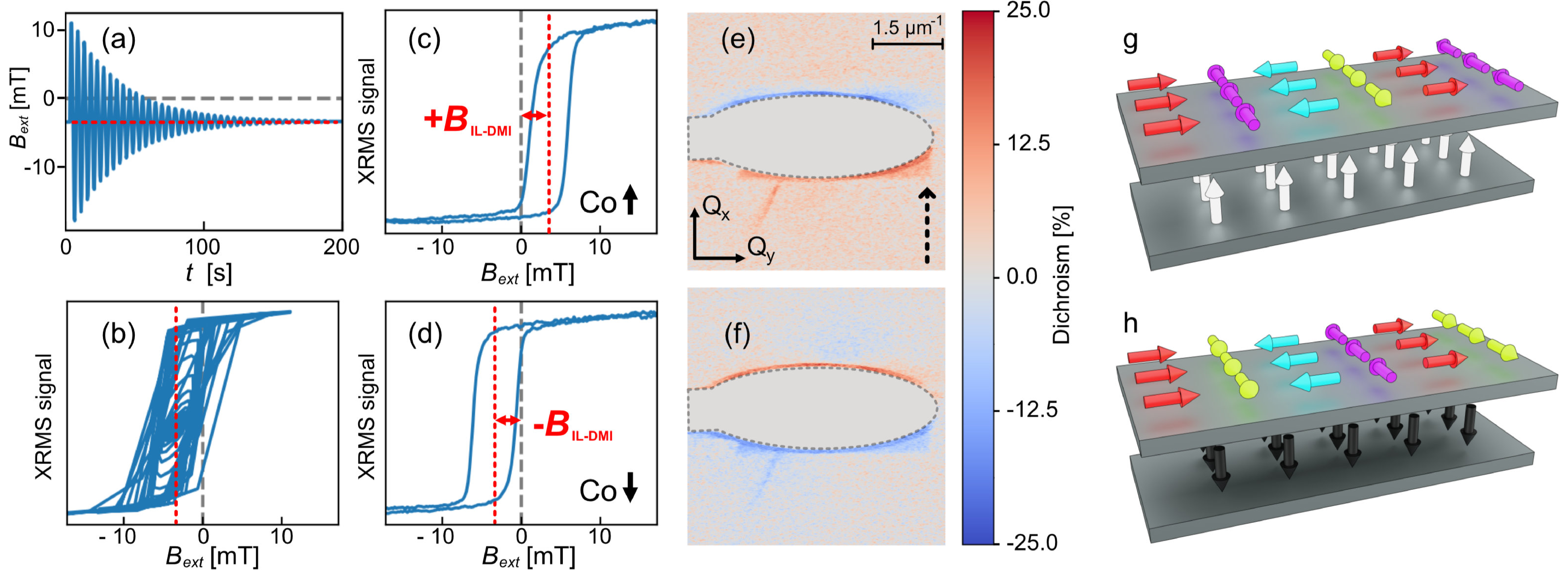}
\caption{\label{fig:figure2} (a) $B_{ext}$ field protocol for demagnetizing the CoFeB layer for Co $\downarrow$, and (b) corresponding XRMS element-specific magnetometry signal during demagnetization. (c,d) XRMS magnetometry signal corresponding to minor IP hysteresis loops of CoFeB layer for (c) Co $\uparrow$ and (d) Co $\downarrow$. The red-dashed line in (a,b,c,d) represents the Co-dependent $B_{\text{IL-DMI}}$. Measurements shown in (b,c,d) are taken at the Co $L_{3}$ edge (778.1 eV). (e,f) Dichroism XRMS maps measured at the Fe $L_{3}$ edge (706.8 eV) for demagnetized CoFeB at (e) Co $\uparrow$ and (f) Co $\downarrow$. In both dichroism XRMS maps, the scattering plane is vertical (black dashed arrow), and aligned with the CoFeB EA. (g,h) Schematic interpretation of the experimental results obtained in maps (e,f), for Co (g) $\uparrow$ and (h) Co $\downarrow$. In both schematics, the color of the IP spins of the top layer denote the configuration of CoFeB, and the OOP spins of the bottom layer denote the Co, using an \textit{hsl} colormap.}
\end{figure*}

\section{Results and discussion}

We focus this work on a "T-SAF", which consists of Si/Ta (4 nm)/Pt (10 nm)/Co (1 nm)/Pt (0.5 nm)/Ru (1 nm)/Pt (0.5 nm)/CoFeB (2 nm)/Pt (2 nm)/Ta (4 nm). The magnetic layers are antiferromagnetically coupled via RKKY interactions, where the Co layer is magnetically hard OOP, and the CoFeB (Co: 60$\%$, Fe: 20$\%$, B: 20$\%$) is magnetically soft IP, with a uniaxial magnetic anisotropy easy axis (EA) along an IP direction ($\hat{x}$).

Prior experiments on this sample \cite{fernandez2019symmetry} demonstrate the existence of IL-DMI, which manifests itself as an exchange bias field acting on the CoFeB layer when performing minor IP hysteresis loops, with the Co layer with net OOP configuration. Due to the chiral interlayer interaction, the sign of the exchange bias flips upon inverting the net OOP orientation of the Co layer.

\subsection{X-ray resonant magnetic scattering}

To investigate the chiral nature of the spin states formed in the SAF, we first perform XRMS experiments at SEXTANTS beamline in SOLEIL synchrotron \cite{sacchi2013sextants}. In these measurements, we record the off-specular scattered signals with a charged-coupled device (CCD) camera to obtain a reciprocal-space-resolved map of the magnetic states present in the system, enabling to identify potential periodicities in the magnetic ordering and length scales \cite{durr1999chiral}. Exploiting circular dichroism in these signals gives direct sensitivity to chiral periodic magnetic textures, which for instance has been used to determine the sign of the intralayer DMI vector in multilayer heterostructures \cite{chauleau2018chirality}.

Prior to recording the XRMS maps, we demagnetize the CoFeB layer by applying an external magnetic field ($B_{ext}$) procedure, as the one in figure \ref{fig:figure2} (a), along the IP EA direction. This way, we reach a state consisting of multiple domains, where their chiral magnetic nature can be studied by this scattering method. To reach such a demagnetized state in the CoFeB layer, we add a constant field offset ($B^{\text{DC}}_{ext}$) to the oscillating component ($B^{\text{AC}}_{ext}$), in order to fully compensate the Co-dependent IL-DMI exchange bias field ($B_{\text{IL-DMI}}$). The resulting magnetometry signal is shown in figure \ref{fig:figure2} (b) for Co $\downarrow$. We do this for OOP initialization field procedures which set the net OOP component of Co $\uparrow$ and $\downarrow$, where the flip in sign of $B_{\text{IL-DMI}}$ is evident in figures \ref{fig:figure2} (c,d).

\begin{figure*}
\centering
\includegraphics[scale=0.195]{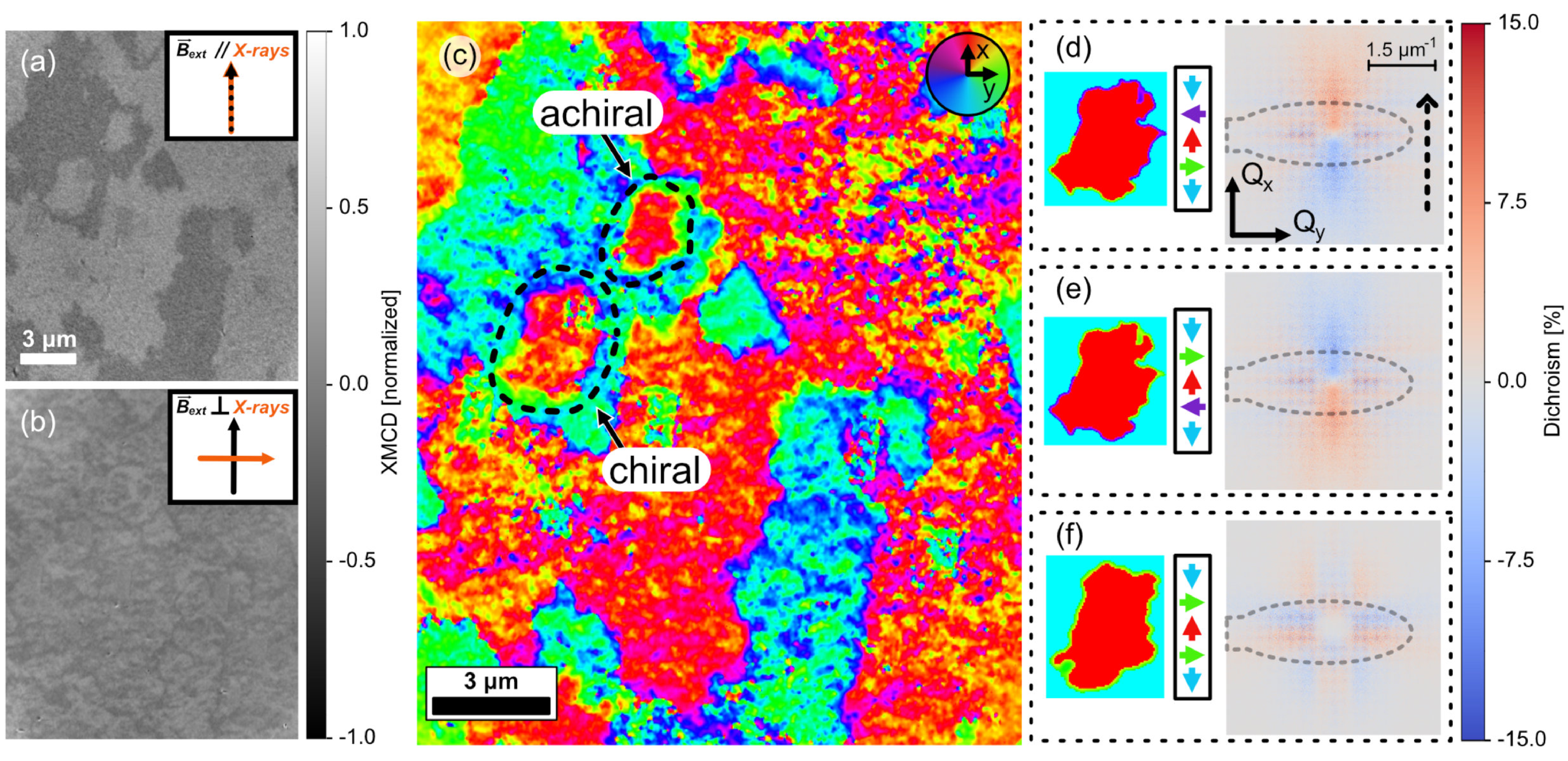}
\caption{\label{fig:figure3} (a,b) XMCD-PEEM images for X-ray beam parallel to the nominal easy (a) and hard (b) axes of the CoFeB layer. The IP projections of $B_{ext}$ (parallel to the EA) and the X-ray beam are given in the inset of each image. (c) Vector reconstruction of the magnetization, where the IP components are represented by the $hsl$ colormap. Two textures, one with vector chirality along the EA direction, and the other with net zero chirality, are circled with a black-dashed line. (d,e,f) Simulated dichroism XRMS maps (right) obtained for configurations (left) of the (d) chiral texture, (e) chiral texture with opposite vector chirality sign after modifying the sign of the DW and (f) achiral texture. As in experiments, the scattering plane direction is represented by the black-dashed arrow and parallel to the EA, and the equivalent beamstop shape from experiments (see figure \ref{fig:figure2}) is outlined as an ellipsoidal dashed line.}
\end{figure*}

After applying $B_{ext}$, we hold $B^{\text{DC}}_{ext}$ constant in order to maintain the demagnetized CoFeB state and record dichroism XRMS maps for both Co orientations. The dichroism XRMS maps are computed via the asymmetry ratio, \textit{i.e.}, a pixel-wise operation $(I_{CL}-I_{CR})/(I_{CL}+I_{CR})$, with $I_{CL}$ and $I_{CR}$ being respectively images taken with circular left and circular right polarized incoming X-rays \cite{chauleau2018chirality}. These are shown in figures \ref{fig:figure2} (e,f) for the X-ray energy set to the Fe $L_{3}$ edge (refer to experimental methods section for further details). The ellipsoidal shape common in both maps (marked by the dashed line) corresponds to a beamstop, used to block the specular reflection whose intensity would otherwise dominate the off-specular signals. Both dichroism maps show asymmetric signals along the scattering plane direction; for Co $\uparrow$, we observe blue-top red-bottom signals, which change to red-top blue-bottom for Co $\downarrow$  (the same behavior is observed at the Co $L_{3}$ edge, refer to supplementary material S1). The dichroic signal peaks exactly at the field required to compensate $B_{\text{IL-DMI}}$, \textit{i.e.}, the sample is fully demagnetized (see supplementary material section S2). When measuring off-resonance, these asymmetric signals disappear, evidencing their magnetic origin (see supplementary material S1). 

The presence of these dichroic asymmetries demonstrates the formation of periodic magnetic textures in the CoFeB layer with a predominant vector chirality along the EA direction \cite{durr1999chiral,mccarter2022structural,kim2022chiral,mccarter2022structural}. In other words, the magnetization vector undergoes 360$^{\circ}$ turns with a specific sense of rotation along this direction. Given the dominating IP anisotropy of the CoFeB layer, the chiral structures corresponding to these signals are IP domains separated by Néel DWs (see section B). From the XRMS signals and the beamstop dimensions, and taking into account the geometry of the beamline, we conclude that the periodicity of these chiral magnetic textures have minimum sizes of $\approx$ 840 nm in real space. However, the exact size distribution for larger lengthscales cannot be retrieved from the maps, as the signal behind the beamstop is not accessible.

The second and more striking conclusion is the relative change in sign of the dichroic signals for both Co OOP directions. This result implies that the sign of the vector chirality in the CoFeB IP layer along the EA is controlled by the net OOP orientation of the Co, as sketched in figures \ref{fig:figure2} (g,h). These results hint to the presence of IL-DMI behind this interlayer symmetry breaking effect, as the other possible source of chirality in CoFeB, \textit{i.e.}, intralayer DMI, is independent on the magnetic state of a different magnetic layer.

\subsection{Photoemission electron microscopy}

To determine the real space configuration of the chiral textures, we perform magnetic imaging of the demagnetized states. For this, we utilize the XMCD-PEEM microscope at the CIRCE beamline \cite{aballe2015alba} of ALBA Synchrotron, exploiting XMCD for magnetic vector sensitivity \cite{foerster2016custom} (refer to experimental methods for further technical details).

The XMCD-PEEM images corresponding to a CoFeB demagnetized state are shown in figures \ref{fig:figure3} (a,b), where we show two orthogonal projections with the X-ray beam parallel to the nominal EA and hard axes (HA) of the CoFeB layer, respectively. These two images are measured at the Co edge, given that the signal-to-noise ratio is significantly better at that edge due to the CoFeB layer's stoichiometry. The signal is expected to come exclusively from the CoFeB layer due to the short mean free path of the outgoing electrons, resulting in a small probing depth. This is evidenced in reference \cite{cascales2023rings}, where the magnetic features are found to be identical at both Fe and Co edges in this SAF.

By combining both projections, we reconstruct the IP components of the magnetization vector \cite{ruiz2018geometrically,ghidini2022xpeem} using the method described in reference \cite{cascales2023rings}. The resulting IP magnetization components are shown in figure \ref{fig:figure3} (c), showing the rough, granular structure of this sputtered multilayer. The reconstructed demagnetized state corresponds to a series of IP domains separated by Néel DWs. In order to understand how these types of textures give rise to the experimental dichroism XRMS maps, we classify them based on their vector spin chirality along the EA direction ($\hat{x}$), which is the property probed in the XRMS experiments. We can thus catalog them in two categories: either chiral, or net achiral. In these two cases, the magnetization vector rotates 360$^{\circ}$ in the plane with either well or not well defined net vector chirality. One example of each type of texture within the reconstruction is labelled and circled with a black-dashed line in figure \ref{fig:figure3} (c).

We now simulate the dichroism XRMS maps that would be produced by the chiral and achiral textures using the off-specular XRMS formalism \cite{flewett2021general}, where the multilayer structure is simplified by modelling it as a single individual ferromagnetic layer, representing the magnetic state of the CoFeB (see supplementary material S3). Prior to the simulations, we apply a series of post-processing algorithms which smoothen and sharpen the DW boundaries in order to reduce noise and remove the granular structure; see figures \ref{fig:figure3} (d,e,f). Furthermore, we have artificially added to the simulations the outline of the real experiment sized beamstop, for direct comparison with experiments.

Figure \ref{fig:figure3} (d) shows the simulated dichroism XRMS maps which would result from the original DW configuration of the chiral texture, showing qualitative agreement with the shape of the experimental signals of figures \ref{fig:figure2} (e,f). Figure \ref{fig:figure3} (e) shows the signals produced by the chiral texture after manually modifying its vector chirality, \textit{i.e.}, the sign of the DW components is reversed, giving opposite dichroic XRMS signal. Differently, the result from the simulation computed with the achiral texture is shown in figure \ref{fig:figure3} (f), evidencing XMCD signals significantly weaker than in figures \ref{fig:figure3} (d,e). Furthermore, the shape of the signal does not match well the experiments. These simulations thus point out the chiral textures observed in XMCD-PEEM as the ones responsible for the experimentally observed signals, as they reproduce the shape, and establish a direct link between the sign of the dichroic XRMS signals and vector chirality of the magnetic textures.

The asymmetric dichroism XRMS maps obtained from experiments, in combination with the conclusions obtained from simulations, show the presence of IP magnetic textures of a dominating vector chirality in the CoFeB (with its sign dependent on the net OOP configuration of the Co layer). However, this interaction is not clearly observed in the XMCD-PEEM magnetization vector reconstruction of figure \ref{fig:figure3} (c). Instead, there is a mixture of IP textures with both chiralities combined with achiral textures. To understand this further, we should consider the significant difference in dimensions of the area probed in both sets of experiments, 30 $\mu$m diameter field of view in XMCD-PEEM, and 300 $\mu$m in XRMS. The XMCD-PEEM data is thus not sufficient to observe the overall symmetry breaking that we probe with XRMS.

From the combination of XRMS and XMCD-PEEM results, we draw the conclusion that in the SAF with IL-DMI under investigation, the overall interlayer scalar spin chirality of the two different ferromagnetic layers, computed across a non-magnetic metallic spacer layer, is preserved. For this, we first consider two neighboring spins along the EA of the CoFeB layer, with a non-zero vector spin chirality. Projecting the resulting vector chirality onto the vector describing the OOP Co spins, yields positive (negative) scalar spin chirality when the vectors align (oppose), demonstrating the preservation of this quantity.

In this work, we solely discuss the states that form when $B^{\text{DC}}_{ext}$ fully compensates $B_{\text{IL-DMI}}$. For a detailed discussion on magnetic states that arise when the CoFeB layer is partially demagnetized, refer to \cite{cascales2023rings}, where we find the emergence of 360$^{\circ}$ DW rings \cite{smith1962noncoherent,heyderman1991360,portier2000formation} favored by $B_{\text{IL-DMI}}$.

\begin{figure*}
\centering

\includegraphics[scale=0.195]{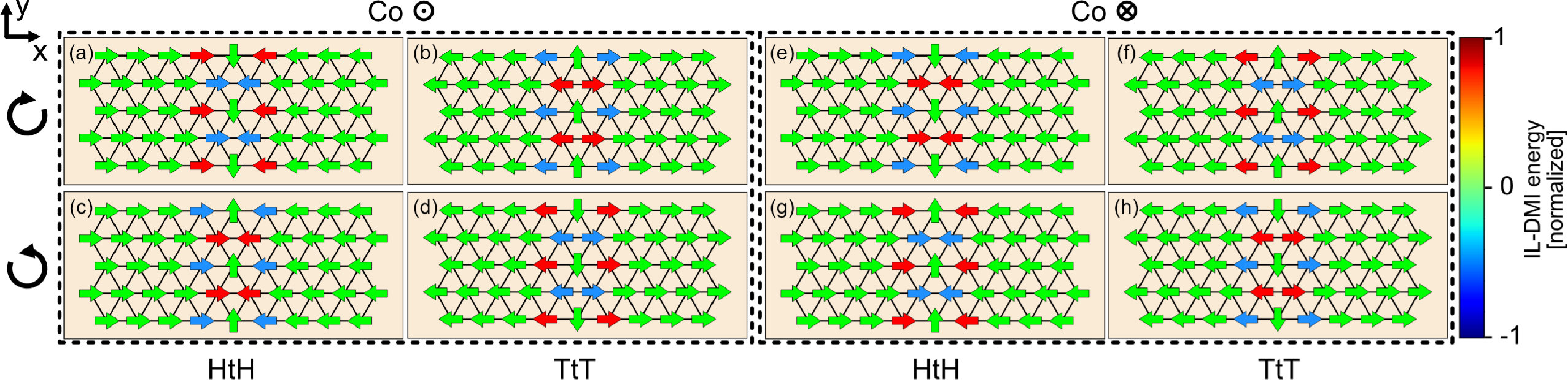}
\caption{\label{fig:figure4} Atomistic IL-DMI energy, with \textit{hcp} stacking, for clockwise 180$^{\circ}$ IP CoFeB (a) head-to-head and (b) tail-to-tail DWs with Co $\uparrow$. Atomistic IL-DMI energy for counter-clockwise 180$^{\circ}$ IP CoFeB (c) head-to-head and (d) tail-to-tail DWs with Co $\uparrow$. (g,h,i,j) Atomistic IL-DMI energies for the same CoFeB configurations as (c,d,e,f), evaluated for Co $\downarrow$. The colormap common to all configurations represents the IL-DMI energy evaluated per spin from the atomistic model.}
\end{figure*}

\subsection{Simulations}

In order to understand the microscopic origin of these results, we perform a set of simulations modelling IL-DMI via two different approaches employed so far in the literature. We first use the method which describes IL-DMI as an additional effective field added to the overall energy due to interlayer symmetry breaking, as in \cite{avci2021chiral,wang2023field,han2019long}. For this, we model the interaction to occur between discretized cells which occupy the same position in the plane (\textit{x,y}), belonging to different layers (different \textit{z} position). In this way, we can evaluate the spatially resolved IL-DMI energy specifying the magnetic configuration of both ferromagnetic layers.

The micromagnetic Hamiltonian associated with the interaction is $H_{\text{IL-DMI}} = \vec{D}_{\text{IL-DMI}}\cdot(\vec{m}_{Co}\times\vec{m}_{CoFeB})$, with $\vec{D}_{\text{IL-DMI}}$ being the overall IL-DMI vector, and $\vec{m}_{Co}$ and $\vec{m}_{CoFeB}$ the normalized magnetization of cells that belong to the Co and CoFeB ferromagnetic layers, respectively. Using vector identities, this expression may be rewritten as $-\vec{m}_{CoFeB}\cdot(\vec{m}_{Co}\times\vec{D}_{\text{IL-DMI}})$, which for uniformly OOP magnetized Co is equivalent to an effective magnetic field acting on the CoFeB with a value equal to $\vec{m}_{Co}\times\vec{D}_{\text{IL-DMI}}$. According to previous observations \cite{fernandez2019symmetry}, we set $\vec{D}_{\text{IL-DMI}}$ parallel to $+\hat{y}$.

For the simulation, we take the magnetic configuration of the chiral texture, left panel of figure \ref{fig:figure3} (d), and compute in a pixel-wise operation the IL-DMI energy for both fully saturated orientations of the Co layer (see supplementary material S4). As expected, the effective field approach does not provide any preferential vector chirality in the CoFeB, thus not explaining our experimental findings.

We now test the atomistic model described in \cite{vedmedenko2019interlayer}, where the IL-DMI energy is obtained by evaluating the Hamiltonian expression $\mathcal{H}_{\text{IL-DMI}} = \vec{D}^{\text{eff}}_{i,j}\cdot(\vec{s}_{i}\times\vec{s}_{j})$. Here, $\vec{D}^{\text{eff}}_{i,j}$ is the effective DMI vector derived utilizing the three-site Levy-Fert model \cite{levy1981anisotropy}. Following the previous approach \cite{vedmedenko2019interlayer}, we assume \textit{hcp} stacking for the layers, with $\vec{s}_{i}$ and $\vec{s}_{j}$ spins corresponding to nearest-neighbor atom pairs located in different layers. The large size of the reconstructed magnetic texture ($\approx$ 3 $\mu$m) makes it unfeasible to evaluate the energy with an atomistic model. We therefore compute the IL-DMI energy for 180$^{\circ}$ DWs representing textures on the CoFeB layer, where the IP anisotropy EA is set along the $\hat{x}$ direction. We do this for all possible combinations of CoFeB vector chirality, \textit{i.e.}, head-to-head (HtH) and tail-to-tail (TtT), and OOP orientation of the Co layer, as summarized in figures \ref{fig:figure4} (a-h).

In all of the maps we observe a "zipper-like" energy distribution; a DW core opposite in energy to the outer part. Thus, IL-DMI indeed creates asymmetries in the energy landscape of 180$^{\circ}$ DWs, reversing for opposite Co orientations. However, all configurations shown here are equivalent in terms of energy, as they exhibit zero overall IL-DMI energy. Furthermore, for a given Co orientation there is no direct link between vector chirality and IL-DMI energy distribution, \textit{e.g.}, figures \ref{fig:figure4} (a,b) show opposite energy distributions for TtT and HtH DWs of CW chirality, which should be equivalent according to experiments. The energy for these simulations is in fact only dependent on the sign of the $\hat{y}$ component of the DW's central spins, regardless of the DW vector chirality. Similar results and conclusions are obtained for different relative orientations of the \textit{hcp} hexagon and anisotropy EA (see supplementary material S5). Thus, the atomistic description of the IL-DMI effect does not explain the experimental results either.

Apart from these two models, we have assessed further options. For instance, a three-spin three-site Hamiltonian is not suitable, as it violates time-reversal symmetry \cite{dos2021proper}. Higher order chiral exchange terms, \textit{e.g.}, a chiral biquadratic interaction \cite{brinker2019chiral}, does not explain our findings either. That is also the case for different additional mechanisms: inhomogeneus RKKY profiles for coupling both ferromagnetic layers, effect of misalignment between external magnetic and IL-DMI effective fields, small IP fluctuations on the OOP Co layer configuration when evaluating energies, dipolar interactions between the layers, or frozen spins. All these possibilities do not currently provide a satisfactory explanation of our experimental results. Thus, further works involving theoretical description and structural configurations of SAFs with IL-DMI are required to understand the complex chiral spin coupling reported here.

\section{Conclusions}

In this work, we study SAFs with IL-DMI using X-ray techniques, in order to investigate what type of chiral magnetic states this interaction promotes. We employ dichroism XRMS in combination with XMCD-PEEM to study the formation of chiral DW configurations upon demagnetizing the CoFeB layer. We discover a direct relationship between vector spin chirality in the CoFeB layer and the net OOP configuration of the Co layer. This mechanism to tune the vector spin chirality by external magnetic fields is different from those utilized until now. For example, through hydrogen absorption \cite{yang2020reversible,chen2021observation}, strain \cite{deger2020strain}, application of electric fields \cite{yang2018controlling}, or electrical currents \cite{kammerbauer2023controlling}. The two main accepted models for IL-DMI fail in explaining this manifestation of chiral interlayer interactions. The same happens for alternative mechanisms which could be present in such multilayered systems. We therefore expect these results will stimulate further work to elucidate the microscopic origin of this new manifestation of chiral interlayer interactions in SAF multilayers. Our results also imply that the overall scalar spin chirality is preserved across the metallic spacer of the SAF, which could result in currently unexplored emergent interlayer-driven topological transport effects.

The formation of complex chiral spin states across multiple layers via interlayer coupling has a great potential for 3D spintronics. This work thus opens prospects for new architectures based on chiral magnetic multilayer devices, where high storage density and interconnectivity could be realized \cite{lavrijsen2013magnetic,fernandez2016magnetic}. 

\section{Acknowledgments}

The authors acknowledge discussions with C. Donnelly and  R. Belkhou, as well as support in computational methods from C. Abert and S. Koraltan. This work was supported by the European Community under the Horizon 2020 Program, Contract No. 101001290 (3DNANOMAG), and UKRI through an EPSRC studentship, EP/N509668/1 and EP/R513222/1. The raw data supporting the findings of this study will be openly available at a repository from TU Wien.

A.H.-R. acknowledges support by the Spanish MICIN (Refs. MICIN/AEI/
10.13039/501100011033/FEDER, UE under grants PID2019-104604RB and PID2022-136784NB) and by Asturias FICYT (grant AYUD/2021/51185) with the support of FEDER funds. S.R-G. acknowledges the financial support of the Alexander von Humboldt foundation. L.S. acknowledges support from the EPSRC Cambridge NanoDTC EP/L015978/1. C.D. acknowledges funding from the Max Planck Society Lise Meitner Excellence Program. The ALBA Synchrotron is funded by the Ministry of Research and Innovation of Spain, by the Generalitat de Catalunya and by European FEDER funds. S.M. acknowledges support from EPSRC project EP/T006811/1. M.A.N and M.F. acknowledge support from MICIN project PID2021-122980OB-C54. E.Y.V. acknowledges financial support provided by the Deutsche Forschungsgemeinschaft (DFG) via Project No.514141286. For the purpose of open access, the author(s) has applied a Creative Commons Attribution (CC-BY) licence to any Author Accepted Manuscript version arising from this submission.

\newpage

\section{Experimental methods}

\textbf{XRMS measurements.} First, the orientation of the Co layer is set by applying a strong ($\approx$ 300 mT) OOP field that saturates the Co by overcoming its strong perpendicular magnetic anisotropy (PMA) \cite{fernandez2019symmetry}. Second, we track the IP signal of the CoFeB layer when applying the field sequences which leave the CoFeB layer in a demagnetized state. For this, we use the element-specific magnetometry capabilities of the beamline where the total signal is integrated with a point detector, rather than resolved in reciprocal space. We take all magnetometry measurements at 17.9$^{\circ}$ of X-ray beam incidence with respect to the sample surface plane, providing large sensitivity to IP components. We show the measurements at the Co $L_{3}$ edge due to the better signal-to-noise ratio found in comparison with the Fe $L_{3}$ edge, caused by the stoichoimetry of the layer (see supplementary material S1). The signals presented here are understood to show the reversal only from the CoFeB layer, the Co is expected to be insensitive to this range of IP fields due to its magnetic hardness. The possibility of having some softer regions with spins slightly changing their orientation with the field would not explain our experimental findings either. Refer to supplementary material S1 for a more detailed discussion.

For all the dichroism XRMS maps discussed in this work, we take 128 images for each individual circular polarization in order to reduce noise levels by subsequent averaging. The angle of incidence is set to 20.1-20.5$^{\circ}$ of incidence with respect to the sample surface plane, as we obtained the best image quality at this angle. We take these measurements at both Fe and Co $L_{3}$-edges, as shown in the supplementary material S1.

\textbf{XMCD-PEEM measurements.} Prior to recording the XMCD-PEEM images, we perform the same procedure as for the XRMS experiments: we initialize the entire system OOP, and we subsequently demagnetize the CoFeB with an oscillating IP magnetic field. We do these measurements for a single Co OOP orientation ($\uparrow$). 

To obtain the XMCD-PEEM images, we compute the pixel-wise asymmetry ratio between opposite circular helicity images \cite{stohr2006magnetism,cascales2024determination}, obtained from averaging 128 images for each polarization. The XMCD-PEEM images originally host different spatial orientations, given that the sample is rotated about the optical axis in order to change the X-ray beam IP projection onto the sample, with the camera in a fixed position. Thus, the images are rotated and aligned with respect to each other utilizing the post-processing algorithms described in \cite{cascales2024determination}, making the different magnetic features spatially overlap. The incidence angle of the X-ray beam is fixed at 16$^{\circ}$ with respect to the substrate plane, giving large sensitivity to IP components.

From the XMCD-PEEM images shown in the text, the EA projection gives information regarding the orientation of the magnetization within the domains, and the HA projection provides information regarding the magnetic configuration of the domain walls which separate the domains. The contrast in the HA image is considerably weaker than that of the EA, since the net magnetization along the HA direction is small when compared with the EA.

\bibliography{bibliography}% Produces the bibliography via BibTeX.

\end{document}